\documentclass[final,3p,times]{elsarticle}

\usepackage[utf8]{inputenc} 
\usepackage[T1]{fontenc}    
\usepackage{hyperref}       
\usepackage{url}            
\usepackage{booktabs}       
\usepackage{amsfonts}       
\usepackage{nicefrac}       
\usepackage{microtype}      
\usepackage{lipsum}
\usepackage{graphicx}
\usepackage{doi}
\usepackage{amsmath}
\usepackage{amssymb}
\usepackage{mathtools}
\usepackage{algorithm}
\usepackage{float}

\begin{document}

\begin{frontmatter}

\title{Forecasting Probability Distributions of Financial Returns with Deep Neural Networks}

\author{
  Jakub Michańków \\
  TripleSun\\
  Krakow, Poland \\
  \texttt{jakub.michankow@triplesun.net} \\
}

\begin{abstract}
This study evaluates deep neural networks for forecasting probability distributions of financial returns. 1D convolutional neural networks (CNN) and Long Short-Term Memory (LSTM) architectures are used to forecast parameters of three probability distributions: Normal, Student's t, and skewed Student's t. Using custom negative log-likelihood loss functions, distribution parameters are optimized directly. The models are tested on six major equity indices (S\&P 500, BOVESPA, DAX, WIG, Nikkei 225, and KOSPI) using probabilistic evaluation metrics including Log Predictive Score (LPS), Continuous Ranked Probability Score (CRPS), and Probability Integral Transform (PIT). Results show that deep learning models provide accurate distributional forecasts and perform competitively with classical GARCH models for Value-at-Risk estimation. The LSTM with skewed Student's t distribution performs best across multiple evaluation criteria, capturing both heavy tails and asymmetry in financial returns. This work shows that deep neural networks are viable alternatives to traditional econometric models for financial risk assessment and portfolio management.
\end{abstract}


\end{frontmatter}

\section{Introduction}

Financial return forecasting has moved from simple point prediction models to distributional forecasting that captures the full uncertainty structure of market dynamics. Traditional econometric approaches often struggle to model the complex, non-linear relationships and time-varying volatility patterns in financial time series. Deep learning opens new possibilities for capturing these patterns, yet most applications still focus on point forecasts rather than the complete distributional properties needed for risk management.

Distributional forecasting in finance is important because accurate risk assessment requires more than just expected returns. Financial institutions, portfolio managers, and regulatory bodies need complete uncertainty quantification to make informed decisions about capital allocation, hedging strategies, and regulatory compliance. Traditional Value-at-Risk (VaR) and Expected Shortfall (ES) calculations rely on distributional assumptions that may not hold during market stress periods.

This study addresses three research questions: First, do deep neural networks provide accurate forecasts of stock return distributions? Second, can these probabilistic forecasts be used for financial risk assessment? Third, do deep learning models outperform classical econometric approaches such as univariate GARCH models?

The main contribution is the systematic evaluation of deep neural network architectures designed for distributional forecasting of financial returns. Custom loss functions based on negative log-likelihood for Normal, Student's t, and skewed Student's t distributions are developed, allowing direct optimization of distribution parameters. The framework uses both 1D CNN and LSTM architectures, benefiting from their strengths in pattern recognition and sequential modeling.

\section{Literature Review}

Probabilistic deep learning and financial forecasting builds on several research areas. The methodological framework is based on probabilistic deep learning principles established by Murphy \cite{murphyMachineLearningProbabilistic2012, murphyProbabilisticMachineLearning2022, murphyProbabilisticMachineLearning2023}, and Duerr et al. \cite{duerrProbabilisticDeepLearning2020}, which provide theoretical foundations for uncertainty quantification in machine learning models.

In energy markets, research by Nowotarski and Weron \cite{nowotarskiRecentAdvancesElectricity2018}, Marcjasz et al. \cite{MARCJASZ2023106843} or \cite{chen2025probabilistic} shows the effectiveness of distributional neural networks for electricity price forecasting, providing precedents for financial markets application.

The evaluation methodology uses established scoring rules for probabilistic forecasts, particularly the framework developed by Gneiting and Raftery \cite{gneiting2007strictly}. This foundation ensures the evaluation metrics are both theoretically sound and practically relevant for comparing forecasting models.

Related research in distributional modeling includes dynamic approaches by Patton et al. \cite{patton2019dynamic} and Thiele et al. \cite{thiele2020modeling}, density combination methods by Opschoor et al. \cite{opschoor2017combining}, and neural network quantile regression by Keilbar and Wang \cite{keilbar2022modelling}. Advanced deep learning approaches include the DeepAR framework \cite{salinasDeepARProbabilisticForecasting2020} and multivariate probabilistic forecasting methods \cite{toubeauDeepLearningBasedMultivariate2019}.

Forecast evaluation methodologies follow frameworks developed by Jordan et al. \cite{jordanEvaluatingProbabilisticForecasts2018}, temporal neural network approaches by Chen et al. \cite{chenProbabilisticForecastingTemporal2020}, and spline-based methods by Gasthaus et al. \cite{gasthausProbabilisticForecastingSpline2019}.

\section{Methodology}

\subsection{Probabilistic Framework}

Financial returns are modeled using a time-varying distributional framework where returns at time $t$ are expressed as:
\begin{equation}
    r_t = \mu(x_t) + \varepsilon_t
    \label{eq:returns}
\end{equation}
\begin{equation}
    \varepsilon_t = \sigma(x_t) z_t, \quad (z_t|x_t) \sim \text{iid } D(\eta(x_t))
    \label{eq:errors}
\end{equation}
\begin{equation}
    \sigma^2_t = \sigma^2(x_t)
    \label{eq:variance}
\end{equation}
where $r_t$ denotes the log return at time $t$, $\mu(x_t)$ is the conditional mean function, $\varepsilon_t$ is the error term, $\sigma(x_t)$ is the conditional volatility function, $z_t$ is the standardized residual, $x_t$ represents the input feature vector at time $t$, and $D$ denotes the assumed probability distribution type with time-varying parameters $\eta(x_t)$:

\begin{equation}
    f_D(z_t;x_t) = 
    \begin{cases}
        f_N^{(1)}(z_t; 0, 1), & \text{where } D \equiv N\\
        f_{St}^{(1)}(z_t; 0, 1, \nu(x_t)), & \text{where } D \equiv St\\
        f_{sSt}^{(1)}(z_t; 0, 1, \nu(x_t), \xi(x_t)), & \text{where } D \equiv sSt
    \end{cases}
    \label{eq:distributions}
\end{equation}
where $N$ denotes Normal distribution, $St$ denotes Student's t distribution, $sSt$ denotes skewed Student's t distribution, $\nu(x_t)$ is the time-varying degrees of freedom parameter, and $\xi(x_t)$ is the time-varying skewness parameter.

\subsection{Distribution Specifications}

The base two-parameter Normal distribution $X \sim N(\mu, \sigma^2)$ has probability density function:
\begin{equation}
   f(x) = \frac{1}{\sigma\sqrt{2\pi}} e^{-\frac{1}{2}\left(\frac{x-\mu}{\sigma}\right)^2}
   \label{eq:normal_pdf}
\end{equation}
where $\mu$ is the mean (location parameter) and $\sigma^2$ is the variance (scale parameter).

The three-parameter Student's t distribution $X \sim St(\mu, \sigma^2, \nu)$ accommodates heavy tails commonly observed in financial returns:
\begin{equation}
  f_{St}(x) = \frac{\Gamma\left(\frac{\nu+1}{2}\right)}{\sqrt{\sigma^2\nu\pi}\Gamma\left(\frac{\nu}{2}\right)}\left(1 + \frac{(x-\mu)^2}{\nu\sigma^2}\right)^{-\frac{\nu+1}{2}}
   \label{eq:studentt_pdf}
\end{equation}
where $\mu$ is the location parameter, $\sigma^2$ is the scale parameter, $\nu > 0$ is the degrees of freedom parameter, and $\Gamma(\cdot)$ is the Gamma function.

The four-parameter skewed Student's t distribution $X \sim sSt(\mu, \sigma^2, \nu, \xi)$ captures both heavy tails and asymmetry:
\begin{equation}
  f_{sSt}(x|\xi) = \frac{2}{\xi+\xi^{-1}}\left[f_{St}(\xi x)H_-(-x) + f_{St}(\xi^{-1}x)H_+(x)\right]
   \label{eq:skewed_pdf}
\end{equation}
where $\xi > 0$ is the skewness parameter (equal to 1 for symmetric distributions) and $H(\cdot)$ denotes the Heaviside function:
\begin{equation}
  H(x) = \begin{cases}
       0 & \text{for } x<0\\
       1 & \text{for } x\geq0
  \end{cases}
   \label{eq:Heaviside}
\end{equation}

For the skewed Student's t distribution, skewness shifts the mean away from the mode, so the conditional mean and conditional variance are:
\begin{equation}
  E(X) = \phi \sigma, \quad \text{Var}(X) = (\gamma - \phi^2) \sigma^2
  \label{eq:sst_moments}
\end{equation}
where:
\begin{equation}
  \phi = \frac{(\xi^2-\frac{1}{\xi^2}) 2 \nu \Gamma(\frac{1}{2}(\nu + 1))}{(\xi + \frac{1}{\xi})(\nu-1) \Gamma(\frac{1}{2} \nu) \sqrt{\pi \nu}}, \quad
  \gamma = \frac{(\xi^3-\frac{1}{\xi^3})}{(\xi-\frac{1}{\xi})} \frac{\nu}{\nu-2}
  \label{eq:sst_parameters}
\end{equation}

Finally, the general probabilistic forecasting framework is expressed as:
\begin{equation}
   p(r_{t+1}|\Psi_{t};\omega) = 
   \begin{cases}
       f_N(r_{t+1}; \mu_{t+1}, \sigma^2_{t+1}), & \text{for } D \equiv N\\
       f_{St}(r_{t+1}; \mu_{t+1}, \sigma^2_{t+1}, \nu_{t+1}), & \text{for } D \equiv St\\
       f_{sSt}(r_{t+1}; \mu_{t+1}, \sigma^2_{t+1}, \nu_{t+1}, \xi_{t+1}), & \text{for } D \equiv sSt
   \end{cases}
   \label{eq:general_framework}
\end{equation}

\subsection{Loss Function Design}

Custom Negative Log-Likelihood (NLL) loss functions are used as the optimization objective minimized by the model for each distribution. For normal and Student's t distributions, the NLL formulation is relatively straightforward, while for the skewed Student's t distribution, we apply the Fernandez and Steel \cite{trottierMomentsStandardizedFernandez2016}, \cite{fernandezBayesianModelingFat1998} transformation.

The general NLL formulation is:
\begin{equation}
   \text{NLL}(\omega) = - \sum_{t=1}^{n} \ln f_D(r_t; \omega)
   \label{eq:general_nll}
\end{equation}
where $\omega$ represents the model parameters, $n$ is the number of observations, and $f_D(r_t; \omega)$ is the probability density function of distribution $D$ evaluated at return $r_t$.

Parameter estimation follows:
\begin{equation}
   \hat{\omega} = \arg \min_{\omega} \text{NLL}(\omega)
   \label{eq:param_estimation}
\end{equation}

For the Normal distribution, the NLL contains three components: constant term, variance penalty, and squared error term:
\begin{equation}
   \text{NLL}(\omega) = \frac{n}{2} \ln(2\pi) + \frac{1}{2}\sum_{t=1}^n \ln\sigma^2_t + \sum_{t=1}^n  \frac{(r_t - \mu_t)^2}{2\sigma^2_t}             
\label{eq:normal_nll}
\end{equation}

This formulation represents standard maximum likelihood estimation for Gaussian distributions.

For the Student's t distribution, additional elements include Gamma function terms and the degrees of freedom parameter $\nu_t$ to account for heavy tails in financial return distributions:
\begin{equation}
\begin{aligned}
   \text{NLL}(\omega) &= -\sum_{t=1}^n \ln\left(\frac{\Gamma\left(\frac{\nu_t+1}{2}\right)}{\Gamma\left(\frac{\nu_t}{2}\right)}\right) + \frac{1}{2}\sum_{t=1}^n \ln(\nu_t\pi) \\
&\quad + \frac{1}{2}\sum_{t=1}^n\ln\sigma^2_t +\sum_{t=1}^n \frac{\nu_t+1}{2} \ln \left(1 + \frac{(r_t - \mu_t)^2}{\nu_t\sigma^2_t}\right)
   \end{aligned}
   \label{eq:studentt_nll}
\end{equation}

For the skewed Student's t distribution, the most complete formulation incorporates the skewness parameter $\xi_t$, captures asymmetry in financial return distributions, and uses piecewise construction via Heaviside functions:
\begin{equation}
\begin{aligned}
   \text{NLL}(\omega) &= -\sum_{t=1}^n \ln\left(\frac{2}{\xi_t+\xi_t^{-1}}\right)+ \sum_{t=1}^n \ln\sigma_t \\
  &\quad - \ln\left[f_{St}(\xi_t z_t;0,1,\nu_t )H_-(-z_t) + f_{St}(\xi_t^{-1}z_t;0,1,\nu_t)H_+(z_t) \right]
\end{aligned}
   \label{eq:skewed_nll}
\end{equation}
where $z_t = \frac{r_t - \mu_t}{\sigma_t}$ and $H(\cdot)$ is the Heaviside function.

\subsection{Neural Network Architectures}

\begin{figure}[H]
\centering
\includegraphics[width=0.8\textwidth]{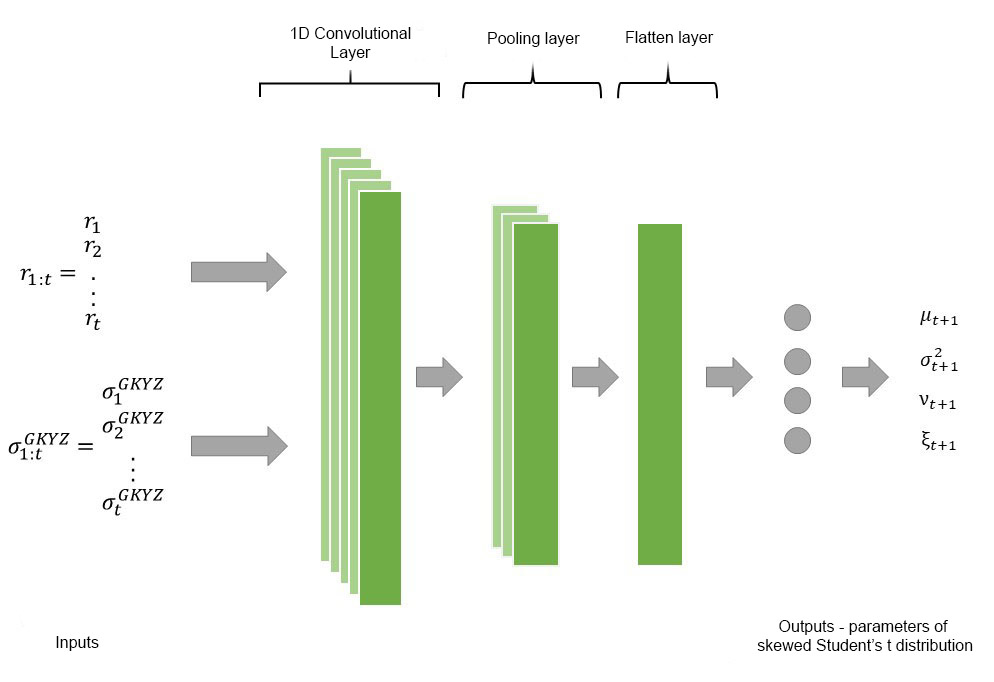}
\vspace{-20pt}
\caption{Architecture of the 1D CNN model for probabilistic forecasting. The model takes return sequences and volatility estimates as input and outputs distribution parameters ($\mu, \sigma^2, \nu, \xi$) for the skewed Student's t distribution. Source: Own study.}
\label{fig:cnn_architecture}
\end{figure}

The 1D CNN model consists of a convolutional layer with 256 filters and kernel size 2, followed by a pooling layer (pool size 2) and flattening layer. The architecture captures local patterns in the return series while maintaining computational efficiency. Figure \ref{fig:cnn_architecture} shows the detailed architecture.

The LSTM model employs three LSTM layers with decreasing neuron counts (128/64/32) to capture long-term dependencies and sequential patterns characteristic of financial time series. The architecture follows the same general structure as shown in Figure \ref{fig:cnn_architecture}, with the convolutional layers replaced by LSTM cells.

Both architectures use dense output layers with neurons equal to the number of distribution parameters being forecasted (2, 3, or 4 parameters).

\subsection{Evaluation Metrics}

The Log Predictive Score (LPS) serves as a basic metric similar to NLL:
\begin{equation}
   \text{LPS} = - \ln p(r_{t+1} | \Psi_t; \theta)
   \label{eq:lps}
\end{equation}
where $\Psi_t$ denotes data used to estimate the predictive distribution at time $t$, and $r_{t+1}$ is the observed value at $t+1$.

The Continuous Ranked Probability Score (CRPS) provides a comprehensive evaluation metric:
\begin{equation}
   \text{CRPS}(F,x) = \int_{-\infty}^{\infty}(F(y) -\mathbf{1}(y\geq x))^2 dy
   \label{eq:crps}
\end{equation}
where $\mathbf{1}$ is the indicator function. Gneiting and Raftery (2007) show it can be rewritten as:
\begin{equation}
   \text{CRPS}(F,x) = E_F |X-x| -\frac{1}{2} E_F|X-X'|
   \label{eq:crps2}
\end{equation}
where $X$ and $X'$ are independent copies of a random variable with CDF $F$. Lower CRPS values indicate better forecasts.

The Probability Integral Transform (PIT) assesses calibration of predictive distributions:
\begin{equation}
   \text{PIT} = F(r_{t+1})
   \label{eq:pit}
\end{equation}
where $F(r_{t+1})$ is the cumulative distribution function evaluated at the observed return $r_{t+1}$. PIT measures the consistency between forecast distribution $F$ and the true distribution $G$. If $F=G$, then it has the uniform distribution $\text{PIT}_{t+1} \sim U(0,1)$.

\subsection{Value-at-Risk and Expected Shortfall}

Value-at-Risk for the long position is defined as:
\begin{equation}
   \text{Pr}\{r_{t+1} \leq \text{VaR}^{l}_{t+1}(\alpha) | \Psi_t \} = \alpha
   \label{eq:var_pr_l}
\end{equation}
\begin{equation}
   \text{VaR}_{t+1}(\alpha) = - r_{t+1|t} - \sigma_{t+1}q^z_\alpha
   \label{eq:var_l}
\end{equation}
where $q^z_\alpha$ denotes the $\alpha$-quantile of the random variable $z_t$.

Expected Shortfall for the long position is calculated as:
\begin{equation}
   \text{ES}^l_{t+1}(\alpha) = E(r_{t+1} | r_{t+1} < \text{VaR}^l_{t+1}(\alpha)) = r_{t+1|t} + \sigma_{t+1} E(z_t | z_t < q^z_\alpha)
   \label{eq:es_l}
\end{equation}

The Kupiec \cite{kupiecTechniquesVerifyingAccuracy1995} (eq. \ref{eq:kupiec_test}) and Christoffersen \cite{christoffersenEvaluatingIntervalForecasts1998} tests evaluate VaR model accuracy:
\begin{equation}
   \text{LR} = 2\left[\ln\left(\left(\frac{N_{p}}{n}\right)^{N_{p}}\left(1-\frac{N_{p}}{n}\right)^{n-N_{p}}\right) - \ln\left(\alpha^{N_{p}}(1-\alpha)^{n-N_{p}}\right)\right]
   \label{eq:kupiec_test}
\end{equation}
where $N_{p}$ denotes the number of VaR exceedances, and $n$ is the number of observations.

The McNeil and Frey \cite{mcneilEstimationTailrelatedRisk2000} test evaluates Expected Shortfall:
\begin{equation}
   U_t = \frac{r_t - \text{ES}^l_{t}(\alpha)}{\hat{\sigma}_t}
   \label{eq:mcneil_test}
\end{equation}
for $r_t < \text{VaR}^l_{t}(\alpha)$, where $\text{ES}^l_{t}(\alpha)$ is the estimated ES at tolerance level $\alpha$ for the long position.

\section{Data and Experimental Setup}

The analysis uses daily log returns from six major equity indices representing different global markets: S\&P 500 (United States), BOVESPA (Brazil), DAX (Germany), WIG (Poland), Nikkei 225 (Japan), and KOSPI (South Korea). The dataset spans from January 3, 2000, to December 31, 2021, providing 2,487 forecasts for each index.

For training and prediction, a walk-forward validation/expanding window approach is used. In the first iteration, the model is trained on data (equal to the training set length, with the validation set comprising 33\% of the training data) and then used for predictions over the next period (equal to the test set length). After that, the window is expanded by additional data and the model is retrained. A single period is predicted each time, based on the last 10 (sequence length) values. Each iteration is trained for 300 epochs. A model checkpoint callback function is used to store the best weights (parameters) of the model based on the lowest loss function value in a specific epoch. The weights are then used for prediction.

Hyperparameter optimization uses both manual tuning and KerasTuner for systematic parameter selection. Table \ref{tab:hyperparameter_tuning} presents the complete hyperparameter configuration.

\begin{table}[H]
\centering
\caption{Model hyperparameters and optimization approach.}
\label{tab:hyperparameter_tuning}
\begin{tabular}{lccc}
  \toprule
  Hyperparameter & Tested values & Selected value & Tool \\
\midrule
Hidden layers & 1-5 &  3 & Manual\\
Neurons per layer & 8-700 & 128/64/32 & KerasTuner\\
Dropout & 0-0.5 & 0.02 & KerasTuner\\
$\ell_2$ regularization & 0-0.5  & 0.002  & KerasTuner\\
Optimizer & Adam/RMSProp/SGD & Adam & Manual\\
Learning rate & 0.0001-0.5 & 0.002& KerasTuner\\
Sequence length & 1-200 & 10 & Manual\\
Window length & 252-Exp./21-1008 & Exp.(min. 1008)-504 & Manual\\
Batch size & 1-Exp. & 128 & Manual\\
Epochs & 10-1000 & 300 (ES) & Manual\\
Kernel size* & 1-5 & 2 & Manual\\
Filters* & 10-512 & 256 & KerasTuner\\
Pool size* & 1-5 & 2 & Manual\\
\bottomrule
\end{tabular}
\normalsize
\end{table}
\vspace{-10pt}
{\footnotesize Note: Exp. denotes expanding window, ES denotes early stopping. * applies to CNN networks. Source: Own study.}

\section{Results}

\subsection{Distributional Forecast Evaluation}

Table \ref{tab:forecast_evaluation} presents the evaluation of model-distribution combinations across selected indices. The LSTM-SSTD (skewed Student's t) configuration consistently achieves the lowest LPS and CRPS values across most indices, indicating better distributional forecast accuracy.

\begin{table}[H]
\centering
\caption{Evaluation of distributional forecasts across indices and model configurations. Bold values indicate best performance for each criterion.}
\label{tab:forecast_evaluation}
\small
\begin{tabular}{lcccccc}
\toprule
\textbf{Index/Model} & \textbf{CNN-N} & \textbf{CNN-STD} & \textbf{CNN-SSTD} & \textbf{LSTM-N} & \textbf{LSTM-STD} & \textbf{LSTM-SSTD} \\
\midrule
\multicolumn{7}{c}{\textit{S\&P 500}} \\
LPS & 1.2820 & 1.2416 & 1.2220 & 1.2632 & 1.2104 & \textbf{1.1933} \\
CRPS & 0.5229 & 0.5248 & 0.5197 & 0.5146 & 0.5137 & \textbf{0.5094} \\
PIT p-value & 2.41e-07 & 1.55e-05 & 0.1144 & 2.41e-07 & 0.0057 & 0.0309 \\
\midrule
\multicolumn{7}{c}{\textit{Nikkei 225}} \\
LPS & 1.6136 & 1.5990 & 1.5865 & 1.6124 & 1.5870 & \textbf{1.5854} \\
CRPS & 0.6914 & 0.6995 & 0.6963 & 0.6892 & 0.6894 & \textbf{0.6874} \\
PIT p-value & 2.41e-07 & 4.41e-05 & 0.0476 & 2.41e-07 & 2.41e-07 & 2.41e-07 \\
\midrule
\multicolumn{7}{c}{\textit{KOSPI}} \\
LPS & 1.3349 & 1.3147 & 1.3172 & 1.3240 & 1.2961 & \textbf{1.2847} \\
CRPS & 0.5285 & 0.5297 & 0.5302 & 0.5246 & 0.5201 & \textbf{0.5165} \\
PIT p-value & 2.41e-07 & 2.41e-07 & 2.41e-07 & 2.41e-07 & 4.70e-07 & 5.08e-06 \\
\bottomrule
\end{tabular}
\end{table}

For the S\&P 500, LSTM-SSTD achieves an LPS of 1.1933 and CRPS of 0.5094, representing the best performance across all metrics. Similarly strong performance is observed for the Nikkei 225 (LPS: 1.5854, CRPS: 0.6874) and KOSPI (LPS: 1.2847, CRPS: 0.5165).

\begin{figure}[H]
\centering
\includegraphics[width=\textwidth]{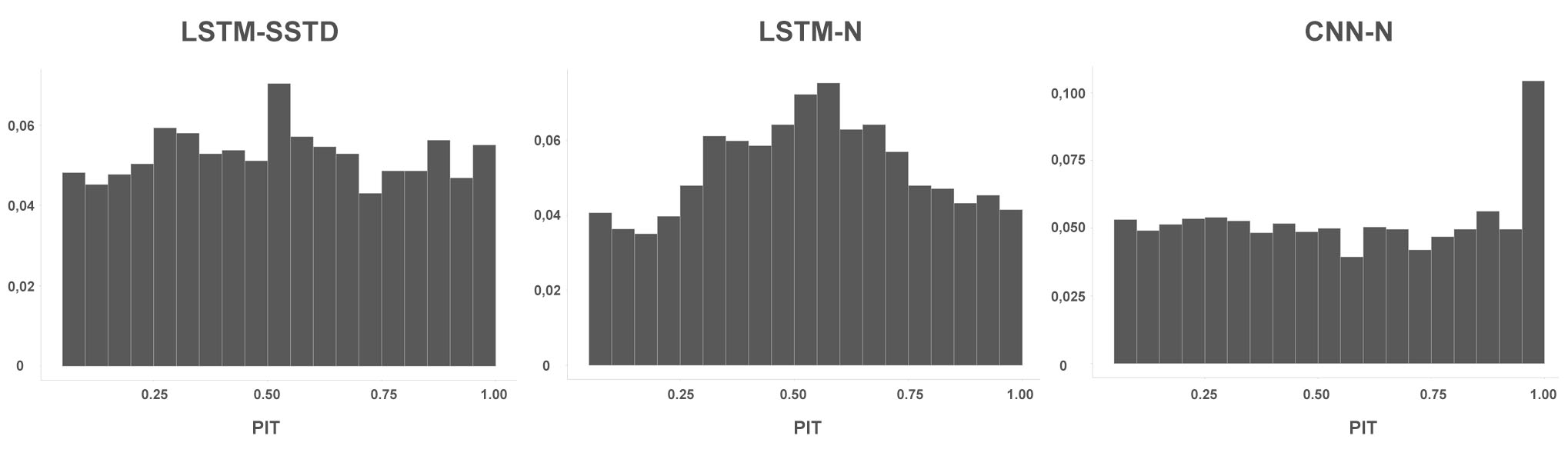}
\vspace{-25pt}
\caption{Distribution plots of PIT values for DAX index across different model configurations. The LSTM-SSTD model (left) shows the best approximation to uniform distribution, indicating better calibration compared to LSTM-N (center) and CNN-N (right) models.}
\label{fig:pit_plots}
\end{figure}

The PIT test results show important information about forecast calibration. The skewed Student's t specifications generally show better calibration, with several configurations achieving p-values above conventional significance thresholds, indicating proper distributional coverage. Figure \ref{fig:pit_plots} provides detailed visualization of these calibration properties.

Figure \ref{fig:pit_plots} shows the calibration quality through PIT histograms for the DAX index. Well-calibrated models should produce PIT values that approximate a uniform distribution. The LSTM with skewed Student's t distribution shows the most uniform PIT distribution, confirming its better calibration properties.

\subsection{Value-at-Risk Performance}

The VaR analysis demonstrates the practical utility of the probabilistic forecasts for risk management. Table \ref{tab:var_exceedances} presents detailed VaR exceedance rates across all model-distribution combinations for each index. The table shows both 5\% and 1\% VaR levels, with statistical test results indicated. 

\begin{table}[H]
\centering
\caption{Percentage shares of VaR(0.05)/VaR(0.01) estimate exceedances across indices and models.}
\label{tab:var_exceedances}
\small
\begin{tabular}{lccc}
\toprule
\textbf{Index/Model} & \textbf{CNN-N} & \textbf{CNN-STD} & \textbf{CNN-SSTD} \\
\midrule
S\&P & \underline{4.50}/1.56* & \underline{4.58}/\underline{0.92}* & \underline{5.30}/\underline{0.88}* \\
NKX & \underline{4.30}*/\underline{1.48} & 4.02*/\underline{0.92}* & 3.53/0.40 \\
DAX & 6.35/2.53 & \underline{\textbf{5.42}}*/\underline{\textbf{1.12}}* & \underline{4.14}*/\underline{0.80}* \\
WIG & \underline{5.07}/1.97 & \underline{4.58}*/\underline{\textbf{1.12}} & 3.69/\underline{0.80}* \\
KOSPI & \underline{5.70}/2.09 & 3.86/\underline{0.56}* & 2.29/0.32 \\
BVP & \underline{5.38}*/\underline{\textbf{1.20}}* & 3.98/\underline{0.72}* & 2.61/0.28 \\
\midrule
& \textbf{LSTM-N} & \textbf{LSTM-STD} & \textbf{LSTM-SSTD} \\
\midrule
S\&P & \underline{\textbf{4.86}}*/1.97 & \underline{5.34}/\underline{\textbf{1.01}}* & \underline{4.50}/\underline{0.84}* \\
NKX & \underline{4.46}*/1.88 & \underline{\textbf{4.58}}*/\underline{\textbf{1.01}}* & 3.81/\underline{0.64} \\
DAX & \underline{5.87}/1.97 & 6.59/\underline{1.16}* & \underline{5.66}/\underline{\textbf{0.88}}* \\
WIG & \underline{4.82}/1.84 & \underline{\textbf{4.95}}/\underline{1.20} & \underline{4.26}/\underline{0.76}* \\
KOSPI & \underline{\textbf{5.42}}*/1.93 & \underline{4.46}*/\underline{\textbf{0.84}}* & 3.37/0.44 \\
BVP & \underline{\textbf{5.18}}*/\underline{1.24}* & 3.65/\underline{0.68}* & 2.41/0.32 \\
\bottomrule
\end{tabular}
\end{table}
\vspace{-10pt}
{\footnotesize Note: Bolded = closest to theoretical tolerance, underlined = correct Kupiec test, * = correct Christoffersen test. Expected exceedances: 124 (5\%) and 24 (1\%) out of 2,487 forecasts. Source: Own study.}

For 5\% VaR, the neural network models achieve exceedance rates very close to the theoretical 5\% level. Notable performances include LSTM-N for S\&P 500 (4.86\%), LSTM-STD for Nikkei 225 (4.58\%), and CNN-STD for DAX (5.42\%). These results satisfy both Kupiec and Christoffersen test requirements in most cases.

The 1\% VaR results show similarly strong performance, with exceedance rates clustering around the theoretical 1\% level. LSTM-STD consistently performs well across multiple indices, achieving 1.01\% for both S\&P 500 and Nikkei 225.

\begin{figure}[H]
\centering
\includegraphics[width=\textwidth]{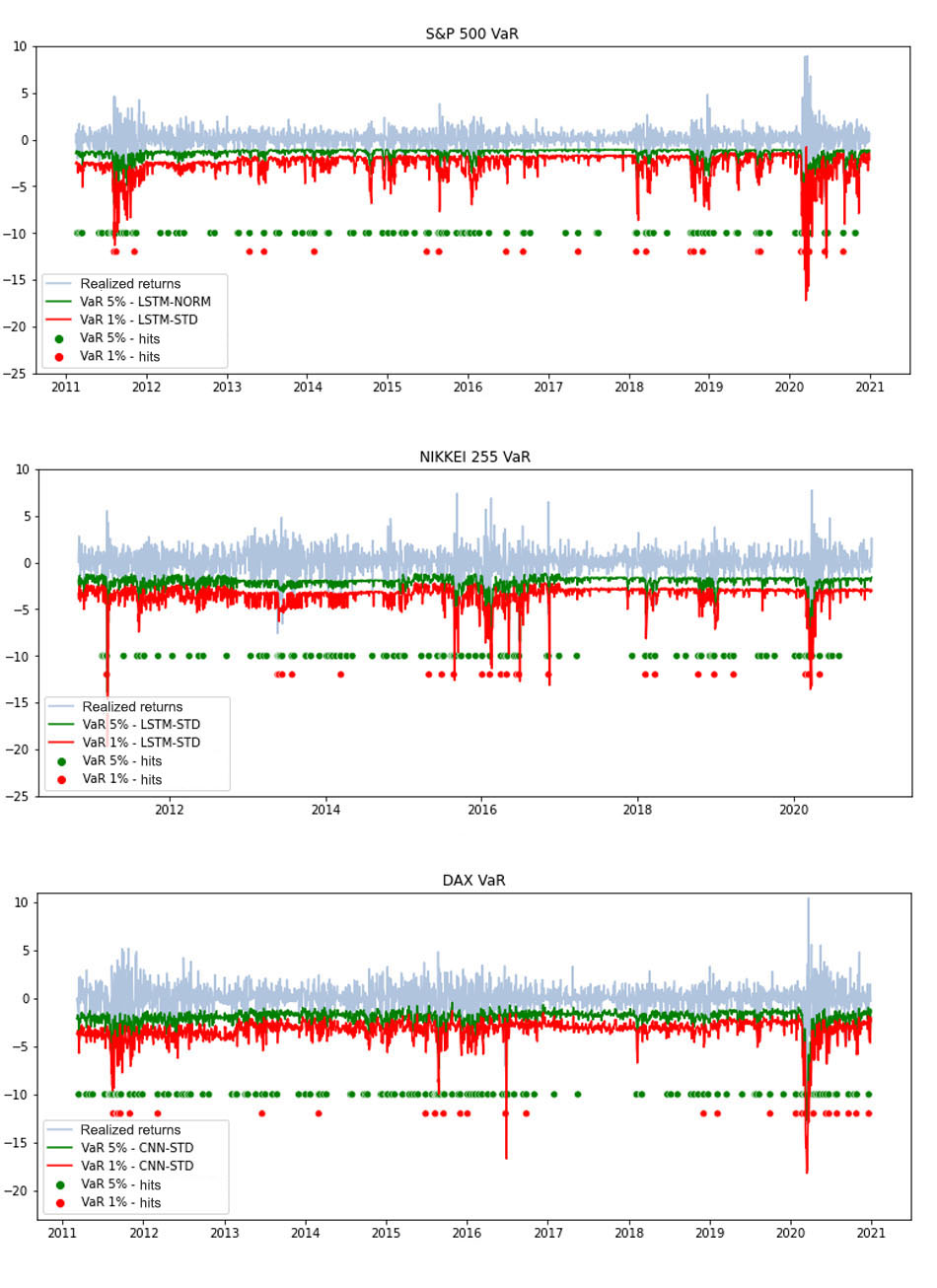}
\vspace{-20pt}
\caption{VaR forecasts for S\&P 500, Nikkei 225, and DAX indices. Blue lines show actual returns, green and red lines represent 5\% and 1\% VaR estimates respectively, with dots indicating VaR exceedances. The models effectively capture periods of high volatility, including the 2020 COVID-19 market stress. Source: Own study.}
\label{fig:var_results_1}
\end{figure}

\begin{figure}[H]
\centering
\includegraphics[width=\textwidth]{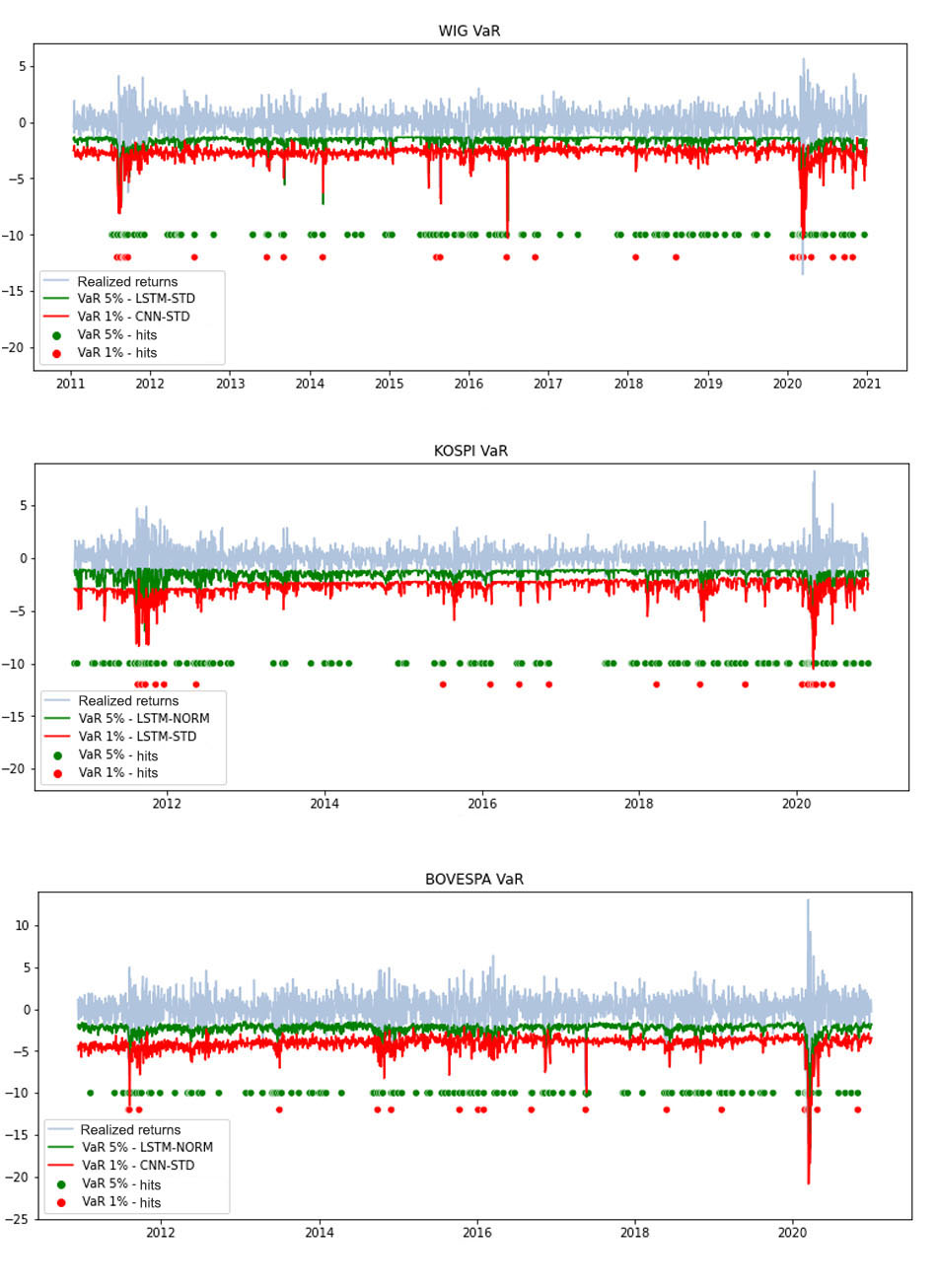}
\vspace{-20pt}
\caption{VaR forecasts for WIG, KOSPI, and BOVESPA indices. The models demonstrate consistent performance across different market conditions and geographical regions, with appropriate clustering of exceedances during crisis periods. Source: Own study.}
\label{fig:var_results_2}
\end{figure}

Figures \ref{fig:var_results_1} and \ref{fig:var_results_2} show the VaR forecasting performance across all six indices. The models successfully capture major market stress periods, with VaR exceedances clustered during times of high volatility such as the 2008 financial crisis, European debt crisis, and COVID-19 pandemic.

Finally, table \ref{tab:vol_prob_snp} shows detailed results, for LSTM model with S\&P500 index, using all three probability distributions. 

\begin{table}[H]
\centering
\caption{Probabilistic forecasts evaluation, S\&P500, LSTM model}
\label{tab:vol_prob_snp}

\begin{tabular}{lccc}
    \toprule
  \textbf{Model/Metrics} & \textbf{LSTM(N)} & \textbf{LSTM(STD)} & \textbf{LSTM(SSTD)} \\
  \midrule
  LPS  & 1,263296 & 1,210462 &  1,193376 \\
  CRPS  & 0,5146218 & 0,5137579 &  0,5094981 \\
  PIT p-value  & 2,412545e-07  & 0,005700126 &  0,03094408 \\
  VaR exc.  & 121/49  & 133/25 & 112/21 \\
  VaR exc. (\%)  & 4,86/1,97 & 5,34/1,005 &  4,50/0,84  \\
  Kupiec 5\%  & 0,7569064 & 0,4310731 & 0,2481387 \\
  Kupiec 1\%  & 1,756379e-05(R) & 0,9791164 &  0,4229126\\
  Christoff. 5\%  &0,4313099 & 0,01607378(R) & 0,0264911(R)   \\
  Chrisotff. 1\% & 1,7719e-08(R) & 0,5192243 &  0,2833833 \\
  ES bootstrap 5\%  & 1,266502e-05 & 0,597149 &  0,5066612\\
  ES sample 5\%  & 1,520095e-07(R) & 0,6093057 & 0,4703062 \\
  ES bootstrap 1\% & --- & 0,7041345 & 0,7884135  \\
  ES sample 1\% & --- & 0,7506336 & 0,8679756 \\
  
  \bottomrule
\end{tabular}
\\
\footnotesize Source: Own study.
\end{table}

The results show superior performance of the skewed Student's t specification, which achieves lowest LPS and CRPS values, indicating better distributional forecast accuracy. The PIT test shows improved calibration as distributional complexity increases, with p-values rising from near-zero for the normal distribution to 0.031 for the skewed Student's t. VaR performance is competitive across all specifications, with exceedance rates close to theoretical levels and most configurations passing the Kupiec and Christoffersen tests. Expected Shortfall tests generally validate the adequacy of the Student's t and skewed Student's t specifications, while the normal distribution shows some rejections.

\subsection{Comparison with GARCH Models}

The comparison includes various GARCH model variants: standard GARCH (G), asymmetric power GARCH (AP), exponential GARCH (E), and GJR-GARCH (GJR), each tested with Normal (N), Student’s t (STD), and skewed Student’s t (SSTD) distributions. For details, see Teräsvirta \cite{terasvirta2009univariate}.

\begin{table}[H]
\centering
\caption{VaR exceedance rates (\%) for neural network vs. GARCH models. Expected exceedances: 5\% and 1\%.}
\label{tab:var_performance}
\small
\begin{tabular}{lcccc}
\toprule
\textbf{Index/VaR} & \textbf{Best NN Model} & \textbf{Exceedances (\%)} & \textbf{Best GARCH} & \textbf{Exceedances (\%)} \\
\midrule
S\&P 5\%   & LSTM-N    & 4.86 & G(STD)   & 5.11 \\
Nikkei 5\% & LSTM-STD  & 4.58 & AP(SSTD) & 4.91 \\
DAX 5\%    & CNN-STD   & 5.42 & E(SSTD)  & 6.03 \\
KOSPI 5\%  & LSTM-N    & 5.42 & AP(SSTD) & 6.15 \\
\midrule
S\&P 1\%   & LSTM-STD  & 1.01 & G(SSTD)  & 0.97 \\
Nikkei 1\% & LSTM-STD  & 1.01 & AP(N)    & 0.92 \\
DAX 1\%    & CNN-STD   & 1.12 & AP(N)    & 0.84 \\
KOSPI 1\%  & LSTM-STD  & 0.84 & AP(N)    & 0.80 \\
\bottomrule
\end{tabular}
\end{table}
\vspace{-10pt}
{\footnotesize
Note: G = GARCH, AP = Asymmetric Power GARCH, E = Exponential GARCH, GJR = GJR-GARCH.
Distributions: N = Normal, STD = Student’s t, SSTD = Skewed Student’s t. Source: Own study.
}

Table \ref{tab:var_performance} compares the best-performing neural networks against classical GARCH specifications. For 5\% VaR, the neural network models achieve exceedance rates very close to the theoretical 5\% level. 

Notable performances include LSTM-N for S\&P 500 (4.86\%), LSTM-STD for Nikkei 225 (4.58\%), and CNN-STD for DAX (5.42\%). These results satisfy both Kupiec and Christoffersen test requirements in most cases. 

The 1\% VaR results show similarly strong performance, with exceedance rates clustering around the theoretical 1\% level. LSTM-STD consistently performs well across multiple indices, achieving 1.01\% for both S\&P 500 and Nikkei 225.

\section{Conclusion}

This study shows that deep neural networks are viable alternatives to traditional econometric models for financial return distribution forecasting. The evaluation across six global markets shows that neural network architectures, particularly LSTM models with skewed Student's t distributions, can provide accurate and well calibrated probabilistic forecasts.

The LSTM architecture's consistent outperformance of CNN models suggests that the sequential nature of financial time series benefits from its memory mechanisms. However, the CNN models remain close, offering possible computational advantages for high-frequency applications.

The practical use for risk management applications is confirmed through VaR testing, with performance competitive to or better than classical GARCH models. The framework's flexibility in capturing complex, non-linear patterns makes it valuable for modern portfolio management and regulatory compliance.

Limitations of this approach include the computational complexity relative to traditional econometric models and the potential for overfitting in volatile market conditions. 

Future research directions include testing additional asset classes, evaluating advanced architectures such as Transformers and xLSTM, conducting sensitivity analysis, investigating quantile-based forecasting approaches, and exploring trading strategy development based on the forecasted distributions.

\subsection*{Source code}
\label{sourcecode}

The complete code can be found and downloaded from Github repository:

\url{https://github.com/jmichankow/deep_learning_probability}


\end{document}